\documentclass[conference]{IEEEtran}
\IEEEoverridecommandlockouts

\usepackage{balance}
\usepackage{amsmath,amsfonts}
\usepackage{algorithm}
\usepackage{algpseudocode}
\usepackage{todonotes}
\usepackage{xcolor,colortbl}

\makeatletter
\@namedef{ver@lineno.sty}{9999/12/31}
\@namedef{opt@lineno.sty}{}
\makeatother
\usepackage[frozencache]{minted}      %
\usepackage{pdftexcmds}                 %

\usepackage{graphicx}

\usepackage{booktabs}
\usepackage{multirow}
\usepackage{hyperref}
\usepackage{url}
\usepackage{framed}
\hypersetup{
    colorlinks,
    citecolor=black,
    filecolor=black,
    linkcolor=black,
    urlcolor=black
}

\newcolumntype{a}{>{\columncolor{lightgray}}r}
\usepackage{colortbl}
\definecolor{lightgray}{gray}{0.9}

\usepackage[backend=biber, abbreviate=true, dateabbrev=true, alldates=year, isbn=false, doi=true, urldate=comp, url=false, maxbibnames=9, maxcitenames=2, backref=false, style=numeric-comp, language=american, giveninits=true, sortcites=true, sorting=none]{biblatex}
\AtEveryBibitem{\clearlist{location}} %
\AtEveryBibitem{\clearfield{series}} %
\AtEveryBibitem{\clearlist{language}} %


\newcommand{\head}[1]{\par\noindent\textbf{#1}\space}

\definecolor{aggreen}{rgb}{0.0, 0.8, 0.6}

\makeatletter%
\begin{document}
\thispagestyle{plain}
\pagestyle{plain}

\makeatletter
\def\ps@IEEEtitlepagestyle{%
  \def\@oddfoot{\mycopyrightnotice}%
  \def\@evenfoot{}%
}
\def\mycopyrightnotice{%
  \hspace*{3mm}\includegraphics[width=2cm]{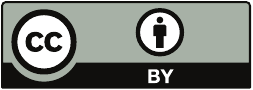}%
  \hspace*{2mm}\raisebox{2.5mm}{%
          \parbox{\columnwidth}{\footnotesize This work is licensed under a Creative Commons \\ Attribution 4.0 International (CC BY 4.0) license.}%
          \hspace*{-68pt}\mbox{\thepage}\hspace{20pt}\fbox{\parbox{.86\columnwidth}{\footnotesize\textsl{Accepted for publication in the 41th IEEE International Conference on Software Maintenance and Evolution (ICSME 2025).}}}%
  }%
  \gdef\mycopyrightnotice{}%
}
\makeatother

\title{The Impact of Fine-tuning Large Language Models on Automated Program Repair}

\newcommand{\SimulaAffiliation}{\affiliation{%
  \institution{Simula Research Laboratory}%
  \city{Oslo}%
  \country{Norway}%
}}

\author{\IEEEauthorblockN{Roman Macháček\IEEEauthorrefmark{1}}
\thanks{\IEEEauthorrefmark{1} Work carried out as a master student at Simula Research Laboratory.}
\IEEEauthorblockA{University of Bern\\
Bern, Switzerland \\
roman.machacek@unibe.ch}
\and
\IEEEauthorblockN{Anastasiia Grishina}
\IEEEauthorblockA{Simula Research Laboratory\\
Oslo, Norway \\
anastasiia@simula.no}
\and
\IEEEauthorblockN{Max Hort}
\IEEEauthorblockA{Simula Research Laboratory\\
Oslo, Norway \\
maxh@simula.no}
\and
\IEEEauthorblockN{Leon Moonen}
\IEEEauthorblockA{Simula Research Laboratory\\
Oslo, Norway \\
leon.moonen@computer.org}
}

\maketitle

\noindent
\begin{abstract}
Automated Program Repair (APR) uses various tools and techniques to help developers achieve functional and error-free code faster. 
In recent years, Large Language Models (LLMs) have gained popularity as components in APR tool chains because of their performance and flexibility.
However, training such models requires a significant amount of resources. 
Fine-tuning techniques have been developed to adapt pre-trained LLMs to specific tasks, such as APR, and enhance their performance at far lower computational costs than training from scratch.

In this study, we empirically investigate the impact of various fine-tuning techniques on the performance of LLMs used for APR. 
Our experiments provide insights into the performance of a selection of state-of-the-art LLMs pre-trained on code. 
The evaluation is done on three popular APR benchmarks (i.e., QuixBugs, Defects4J and HumanEval-Java) and considers six different LLMs with varying parameter sizes (resp. CodeGen, CodeT5, StarCoder, DeepSeekCoder, Bloom, and CodeLlama-2). 
We consider three training regimens: no fine-tuning, full fine-tuning, and parameter-efficient fine-tuning (PEFT) using LoRA and IA3. 
We observe that full fine-tuning techniques decrease the benchmarking performance of various models due to different data distributions and overfitting. 
By using parameter-efficient fine-tuning methods, we restrict models in the amount of trainable parameters and achieve better results.  
\end{abstract}

\begin{IEEEkeywords}
large language models, 
automated program repair,
parameter-efficient fine-tuning,
AI4Code, 
AI4SE, 
ML4SE.
\end{IEEEkeywords}

\section{Introduction}

\noindent
Software development, maintenance, and evolution are expensive processes, both in terms of money and time~\cite{bugimpact}. 
Supporting the efficiency of software engineers responsible for these workflows can save significant resources and enable companies to speed up the production and delivery of products without loss of quality. 
One of the key challenges that software engineers face is the occurrence of software defects, or bugs, which are unintended errors in the code that cause deviations from expected behavior. 
These defects vary in complexity, from simple one-line syntax errors to intricate multi-line logic bugs that can span multiple files and components. 

Automated Program Repair (APR) aims to support developers with the software maintenance and evolution process, 
by helping them to fix any bugs they encounter and achieve their goals faster.   
Many techniques with varying characteristics and performance exist, depending mainly on the type and complexity of the bug~\cite{huang2023:survey}. 
Some methods specialize in specific programming languages~\cite{getafix, drain2021:deepdebug, zhang2024:pydex}, while others use specific patterns found in common bugs~\cite{liu2018mining, liu2019avatar}. 
A relatively new direction in APR methods is based on using Large Language Models (LLMs)~\cite{jiang2023impact, zhang2023survey, fan2023automated}, which have made significant progress in text-based tasks, including, for instance, text summarization, translation, and chatbots. 
When trained on programming languages, such models can learn to translate from buggy to fixed code rather than between languages.

While LLMs that are pre-trained on code already have considerable capabilities, 
fine-tuning them on a specific task such as APR can further improve their performance. 
However, the fine-tuning of LLMs is not without its challenges, 
such as computational overhead due to the large datasets and models, 
and overfitting of the models to the datasets~\cite{kaddour2023:challenges}. 
Nevertheless, LLMs for APR remain a very active research topic due to the quality of the solutions (or patches) they generate compared to those generated by existing APR methods.
One promising strategy to combat the high costs of fine-tuning, %
is the use of parameter efficient fine-tuning (PEFT) techniques~\cite{xu2023:parameterefficient}.
These techniques can freeze parts of the original LLMs and enable fine-tuning on a reduced subset of parameters.
However, a potential disadvantage of limiting the number of trainable parameters is that the model loses power and performance decreases.
The investigation described in this paper aims to better understand the trade-off between efficiency gains and power loss in the context of fine-tuning models for APR. 

\head{Contributions:}
We conduct an extensive empirical study on the impact of fine-tuning LLMs on their APR performance. 
We provide a detailed comparison of different fine-tuning strategies for APR and offer practical insights into leveraging LLMs for automated software maintenance and evolution.
Specifically, we make the following contributions:
\begin{itemize}\renewcommand{\labelitemi}{$\star$}
\item \emph{Establish a baseline of pre-trained LLMs for the program repair task.}
    We select six state-of-the-art LLMs pre-trained on code 
    (resp. CodeGen~\cite{nijkamp2023codegen}, CodeT5~\cite{wang2021codet5}, Bloom~\cite{workshop2023bloom}, CodeLlama-2~\cite{roziere2024:codellama}, StarCoder~\cite{li2023starcoder}, and DeepSeek-Coder~\cite{guo2024:deepseekcoder}),
    and evaluate their out-of-the-box performance on a selection of APR benchmarks without applying any fine-tuning.
    Following earlier work by Jiang et al.~\cite{jiang2023impact}, we use three widely adopted APR benchmarks (i.e., QuixBugs, Defects4J, and HumanEval-Java) and count the number of fixed examples in each benchmark. 
    Finally, to help find the preferred input format for the selected LLMs, 
    we analyze how including or omitting the buggy line(s) affects the LLM's performance on the benchmarks.  
    Our findings provide insights into the suitability of non-fine-tuned models for APR and establish a baseline for the next steps of our study.
\item \emph{Assess the impact of full-model fine-tuning on APR performance.}
    We empirically examine the effects of full-model fine-tuning of the selected LLMs on an APR-specific dataset of bug-fix pairs on their APR capabilities.
    This helps us to understand whether allowing adjustment of all parameters during fine-tuning improves or degrades the repair performance.
    The fine-tuning dataset was collected by Zhu et al.~\cite{datasetimpact} and used earlier in the study by Jiang et al.~\cite{jiang2023impact}. 
    We measure model effectiveness using established APR metrics such as exact match and CodeBLEU~\cite{ren2020:codebleu}, along with model loss on the training and validation sets, and make a final evaluation of the model performance on the three APR benchmarks.
    Since the datasets for fine-tuning and benchmarking do not have the same bug-fix distribution, this simulates scenarios closer to the real world.
	
\item \emph{Explore the effects of parameter-efficient fine-tuning (PEFT) on APR performance.}
    We examine adapter-based PEFT techniques, focusing on LoRA~\cite{lora} and IA3~\cite{ia3}, 
    which optimize training efficiency by fine-tuning fewer parameters.
    We compare these approaches to full fine-tuning, analyzing trade-offs between computational efficiency and performance.
    Our results highlight the potential of overcoming resource constraints using PEFT.

\item \emph{Analyze the impact of LoRA hyperparameters on APR performance.}
    We systematically vary LoRA hyperparameters to assess their impact on APR performance.
    This part of the study provides a deeper understanding of how parameter-efficient fine-tuning configurations affect LLM effectiveness for APR.
    The insights from this experiment help to optimize PEFT strategies in future APR research.

\item \emph{Provide a comprehensive replication package for experimental setup and evaluation.}\footnote{~Replication package: \url{https://doi.org/10.5281/zenodo.16359186}.} 
    We provide a replication package with code and results for our study to enable others to build on our work. 
    Our code builds on the code provided by Jiang et al.~\cite{jiang2023impact}, 
    adapting it to accommodate the benchmarking of additional models and parameter-efficient fine-tuning. 
    Our extensions were developed in a modular fashion to facilitate the easy integration of additional models, 
    techniques, and benchmarks~\cite{machacek2025:replication}. 
\end{itemize}

\section{Background and Related Work}

\subsection{Large Language Models}

\noindent The use of LLMs has gained a lot of popularity in recent years, due to the increase of computational resources and breakthroughs in deep learning, such as transformer models.
The transformer architecture~\cite{vaswani2017:attention} allows models to efficiently learn long-range dependencies in text data. Contrary to previous approaches towards sequence processing, transformers do not use recurrent sequential structures to process sequences, instead, they process sequences in parallel through multiple layers to produce output sequences, thus improving upon problems faced by RNN-based architectures. A core component of transformers is self-attention, allowing them to focus on different parts of the sequence to capture token dependencies.

Training of LLMs is typically done using semi-supervised learning approaches. For example, during pre-training, we may mask some of the tokens in a sentence and make the model predict the masked or missing token in the sequence. Another approach is to predict the next token in sequence, which is achieved by masking tokens after the predicted token. Currently, the fill-in-the-middle (FIM) approach is used for decoder models, where the whole sentence is used to predict the next token in sequence by re-structuring the input sentence using special tokens~\cite{bavarian2022:efficient}. 

By pre-training LLMs we obtain a model with knowledge about various domains, problems, and grammatical structure. 
Further fine-tuning this model then improves the model's performance and understanding of the specific task at hand, which has been done for a diverse set of software engineering tasks~\cite{zhang2024:appt,yu2024:finetuning,shang2025:largescale,nashaat2024:efficient}.

Nowadays, research and university clusters cannot compete with industry giants in training the biggest LLMs due to scarce resources, thus restraining researchers with limited computational assets from the state-of-the-art models. 
One approach for reducing training time and memory requirements is to use only part of a model's parameters or layers, such as training only last layers of the model, while keeping the rest of the parameters \emph{frozen} to decrease the amount of trainable parameters. 
Another common alternative to freezing is \emph{pruning}~\cite{chen2022understanding}, in which weights close to zero are clipped to zero, thus making them sparse. 

\subsection{Parameter Efficient Fine-tuning}

\noindent
Parameter efficient fine-tuning (PEFT)~\cite{xu2023:parameterefficient} is another technique to address the challenge of having an LLM with a large amount of trainable parameters.
PEFT techniques address this challenge by freezing the whole model and adding an additional trainable structure, called an adapter. 

A potential disadvantage of limiting the number of trainable parameters is that the model loses its power and performance decreases. 
The investigation described in this paper aims to better understand the trade-off between efficiency gains and power loss in the context of fine-tuning models for APR. 

In particular, we investigate the impact of the LoRA~\cite{lora} and IA3~\cite{ia3} adapters on APR because of their promising results in other applications~\cite{haque2025:systematic}. %

\subsubsection{LoRA}
\label{section:lora}
Low-Rank Adaptation (LoRA) is a method developed by Hu et al.~\cite{lora}. The main idea behind LoRA is to decompose large-weight matrices from Large Language Models into smaller low-ranking factors, thus reducing memory and increasing the computational efficiency of the model during training and inference.
The inspiration comes from the observation that pre-trained models with a large number of parameters inherently exhibit a low-dimensional structure~\cite{aghajanyan2020intrinsic}.
Furthermore, pre-trained LLMs learn efficiently even after being projected into a subspace of lower dimension~\cite{lora}.

\begin{figure}[tb]
    \centering
    \includegraphics[width=0.5\linewidth]{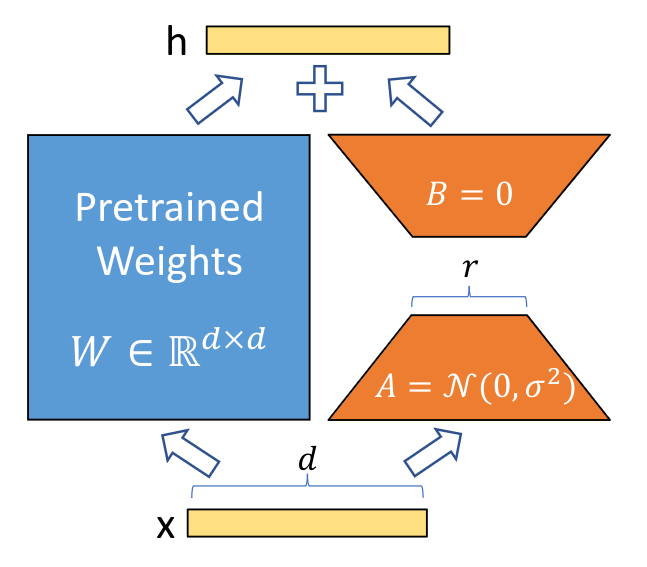}
    \vspace{-7pt}
    \caption{Illustration of LoRA decomposition~\cite{lora}.}
    \label{fig:lora}
\end{figure}

The goal is to project the weight matrix $W_0 \in \mathbb{R}^{d \times k}$ to low-rank decomposition $W_0 + \Delta W = W_0 + BA$ as can be seen in Figure \ref{fig:lora}, where $B \in \mathbb{R}^{d \times r}, A \in \mathbb{R}^{r \times k}$ and the rank $r < \min(d, k)$, resulting in $d \cdot r$ trainable parameters. 

The pre-trained weights are frozen during training, $\Delta W$ is approximated by the parameterized matrices $A, B$. Forward pass, or inference, then corresponds to the use of pre-trained weight $W_0$ together with an approximation of $\Delta W$ during fine-tuning:
$$
f(x) = W_0x + \Delta Wx = W_0x + BAx
$$

LoRA has been adopted in many frameworks and serves as an efficient method of training LLMs. The advantage of LoRA lies in the ease of use for applications of Large Language Models to new domains with limited resources, where instead of fine-tuning the whole model for small problems, we approximate the update on a subspace, saving time and resources.

\subsubsection{IA3}
\label{section:ia3}
\begin{figure}[t]
    \centering
    \includegraphics[width=0.7\linewidth]{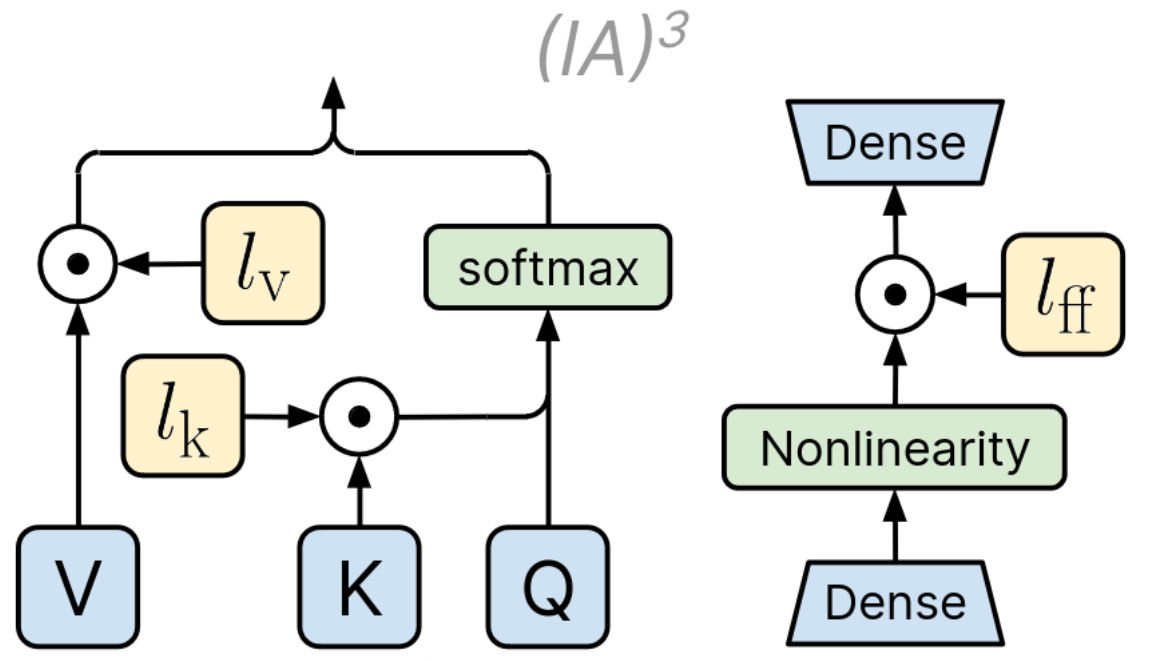}
    \caption{Illustration of IA3 adapter~\cite{ia3}.}
    \label{fig:ia3}
\end{figure}

Infused Adapter by Inhibiting and Amplifying Inner Activations (IA3)~\cite{ia3} is a parameter-efficient fine-tuning technique that promises improvement over LoRA. The idea of IA3 lies in optimizing the transformer architecture directly, more specifically the attention mechanism, by adding three vectors in order to re-scale key and value matrices. Similarly to LoRA, pre-trained weights remain frozen but instead injected scaling vectors are trained, as illustrated in Figure \ref{fig:ia3}. Scaling vectors (as opposed to low-rank matrices) reduce the parameters to save time and resources during training and inference.  Consider the vectors $l_k \in \mathbb{R}^{d_k} , l_v \in \mathbb{R}^{d_v}$, and $ l_{ff} \in \mathbb{R}^{d_{ff}}$. Attention $A$ mechanism of the transformer then becomes~\cite{ia3}:
$$
A = \sigma \left( \dfrac{Q(l_k \cdot K^T)}{\sqrt{d_k}} \right)(l_v \cdot V)
$$

Here, $\sigma$ corresponds to the softmax function. The IA3 method adds $L(d_k + d_v + d_{ff})$ parameters for the $L$-layer encoder part of the transformer and $L \cdot (2d_k + 2d_v + d_{ff})$ for a $L$-layer decoder. The forward pass in the position feed-forward network of the transformer can be written as~\cite{ia3}:
$$
f(x) = l_{ff} \cdot f(W_1x))W_2
$$

Similarly as in LoRA, we can have multiple IA3 adapters on top of a single pre-trained model, each one trained and serving a different task. 
Because the base pre-trained model can be merged into the IA3 adapter, there is no overhead during inference.
According to experiments by Liu et al.~\cite{ia3}, IA3 outperformed full fine-tuning and performs better than LoRA while using fewer trainable parameters. 

Before focusing on program repair in the next section, we briefly review studies that consider PEFT for other software engineering (SE) tasks, such as code generation, clone detection, defect detection, and code summarization. 
\Citeauthor{zhuo2024:astraios} conducted one of the first studies of PEFT for code LLMs and found that while complete fine-tuning generally performs best on code generation and code comprehension benchmarks, LoRA offered the most competitive trade-offs between cost and performance, particularly for the largest (16B parameter) models in their study~\cite{zhuo2024:astraios}.    
\Citeauthor{weyssow2025:exploring} empirically investigated PEFT for code generation and compare it to in-context learning (ICL) and retrieval-augmented generation (RAG) to guide the synthesis with task-specific examples~\cite{weyssow2025:exploring}. 
They found that PEFT performs superior to ICL and RAG across a diverse set of LLMs and three representative Python code generation datasets.
\Citeauthor{afrin2025:resourceefficient}~\cite{afrin2025:resourceefficient} explored the impact of PEFT, in particular Quantized LoRA (QLoRA) on code summarization. 
They found that QLoRA enables efficient fine-tuning of code LLMs for summarization and requires minimal parameter adjustment compared to full model fine-tuning.
Finally, \citeauthor{haque2025:systematic} conducted a recent systematic literature review on PEFT for code LLMs~\cite{haque2025:systematic}.

\subsection{Automated Program Repair}

\noindent
Automated Program Repair (APR)~\cite{zhang2023survey} is a field of software engineering that deals with the identification and repair of bugs. Bugs are flaws, mistakes, or errors, in the source code. APR can be useful for the maintenance of already existing repositories and the development of new projects, increasing code quality, and decreasing the time spent on debugging. To accomplish this goal, several techniques and algorithms were created in order to analyze the code, detect bugs in it, and generate the corresponding patches, thus fixing the bug. 

In the following, we are going to discuss the learning-based APR methods and techniques, surveyed by Huang et al.~\cite{huang2023:survey} and Zhang et al.~\cite{zhang2023survey}.
Their goal is to use machine learning algorithms to learn and repair bugs and leverage huge amounts of software engineering data. 
With the increasing amount of data, this approach gained significant popularity over the last few years and was used to repair errors~\cite{li2022transrepair, yasunaga2021:breakitfixit, 10172803, jiang2023impact} and vulnerabilities~\cite{fu2022:vulrepair, pearce2022examining, islam2024:llmpowered} in code. However, learning-based methods suffer from a number of limitations, including internal interpretation of models and the repair of long sequences~\cite{huang2023:survey}.

Many methods for automatic repair of software bugs were developed, using abstract syntax trees~\cite{dlfix}, code similarity and automatic matching~\cite{dlsimilarities}, convolutional neural networks~\cite{Lutellier2020CoCoNuTCC}, code reviews~\cite{huq2020review4repair}, grammar rules~\cite{grammarbased} along with syntax~\cite{zhang2024:pydex}, code aware beam-search~\cite{jiang2021:cure} and hybrid techniques combining ML techniques with fault localization~\cite{li2022dear}. 
The focus has also been on having models capable of repairing programs in multiple languages~\cite{circle} with multiple lines and locations~\cite{li2022dear}, called multi-hunk~\cite{huang2023:survey}, repair capabilities. Repair of easy programming errors, usually one-line semantic and syntactic errors, utilized RNNs~\cite{pu2016skp}, graph neural networks~\cite{yasunaga2020graphbased}, data perturbation~\cite{yasunaga2021:breakitfixit}, error messages~\cite{repairnet}, transformer-based models~\cite{li2022transrepair}, parameter-efficient methods for code generation~\cite{silva2024:repairllama, weyssow2025:exploring} along with prompt engineering~\cite{paul2023:enhancing}. 

With LLMs, we are able to use datasets and pre-train models to gain an understanding of code syntax and logic before fine-tuning models for a specific code repair task and language.  
By providing models with context, the code around the bug, we allow models to make context-aware repairs. Pre-training models on vast amounts of code alleviates a problem of limited patch variety, since the models do not need to be trained on relatively small datasets consisting of bug-fix pairs. Recently, LLMs were used for APR with~\cite{jiang2023impact} and without fine-tuning~\cite{10172803,sobania2023analysis} and achieved promising results.  
Moreover, there are some recent works which observed the benefit of using PEFT techniques for APR~\cite{yang2024multi,silva2024:repairllama,li2024comprehensive}.
For instance, Silva et al.~\cite{silva2024:repairllama} presented an approach for APR which fine-tuned a
CodeLlama-7b model with LoRA. The performance of LoRA was better than full fine-tuning.

Jiang et al.~\cite{jiang2023impact} carried out APR experiments with full fine-tuning on different LLMs and share a framework. We build upon their framework but use a different set of models and add PEFT methods on top of these.
The closest work to ours is the recent study by Li et al.~\cite{li2024:exploring}, who also investigated PEFT techniques for APR, and the work by Huang et al.~\cite{huang2025:comprehensive}, who carried out a study on the fine-tuning of LLMs for APR concurrently with our investigation.
In contrast to our experiments, Li et al. considered instruction-tuning datasets while we perform regular fine-tuning. 
We compare our findings where possible, i.e., where the same models and hyperparameters are used, and thus investigate how instruction-based fine-tuning~\cite{li2024:exploring} affects the results compared to NMT-style fine-tuning as is also done by Jiang et al.~\cite{jiang2023impact}.

\section{Experimental Design}

\noindent
In this section, we describe experimental details (e.g., datasets, models, metrics) to answer the following three research questions:

\begin{itemize}
    \item \textbf{RQ1:} What is the benchmarking performance of Large Language Models pre-trained using code on the program repair task, without further fine-tuning?
    \item \textbf{RQ2:} How does full fine-tuning of LLMs on a dataset of bug-fix pairs affect the resulting APR performance?
    \item \textbf{RQ3:} What is the effect of adapter-based parameter-efficient fine-tuning of LLMs on APR benchmarks?
    \item \textbf{RQ4:} How do hyperparameters of LoRA affect the performance of selected LLMs?
\end{itemize}

\subsection{Datasets}
\label{section:datasets}

\noindent
Here we outline the three benchmarking datasets, to evaluate LLMs on the APR task, and one dataset for fine-tuning LLMs.
First, we introduce the three benchmarking datasets: Defects4J, QuixBugs, HumanEval-Java.
Among these three benchmark datasets, Defects4J contains real-world projects with long context, while HumanEval-Java and QuixBugs include smaller programming problems with shorter context.

\textbf{Defects4J}~\cite{defects} is a collection of real-world Java bugs collected from multiple open-source projects. The amount of bugs in the database varies depending on the Defects4J version, in our case we work with Defects4J 2.0.1 consisting of 835 active bugs and 29 deprecated bugs, of which 219 are used. The infrastructure provided contains information about the bug, along with tests and a fixed code. %

\textbf{QuixBugs}~\cite{quixbugs} is a multilingual collection of 40 programs translated into Python and Java collected from the Quixey Challenge, where we use only Java programs. Test cases are provided with an infrastructure for the validation of each of the programs. The defects found in each program are further classified. %

\textbf{HumanEval-Java}~\cite{jiang2023impact} is a manually created dataset based on the HumanEval dataset from which 163 bugs are used. The reason for doing so is to eliminate the threat that models have already seen the dataset during pre-training~\cite{jiang2023impact}. Each of the programs in the HumanEval dataset, together with their test cases, was converted from Python to Java.
HumanEval-Java follows a similar structure as QuixBugs, containing both one-line and multi-line bugs together with the fixed program files and their test cases. %

\textbf{Finetuning:} For fine-tuning our LLMs, we use the same dataset as Jiang et al.~\cite{jiang2023impact}, introduced by Zhu et al.~\cite{datasetimpact}.
As the dataset has no formal name, we refer to it as \textbf{CLM}, based on the GitHub repository name for the work of Jiang et al.~\cite{jiang2023impact}, where we obtained it.\footnote{~\url{https://github.com/lin-tan/clm}}
The CLM dataset was created from Java projects on GitHub, consisting of 1,083,185 commits from March 2011 to March 2018. Filtering of commits was done to obtain only those patches that correspond to either an update or an insertion of a single-hunk statement, in line with the three test datasets. 
Moreover, to avoid benchmark contamination through data leakage, all patches related to projects in Defects4J were removed based on a comparison using abstract syntax trees~\cite{datasetimpact}.
The resulting dataset contains 143,666 instances, of which 129,300 (80\%) are in the training dataset and 14,366 (20\%) in the test dataset. %
By using this cleaned CLM dataset for fine-tuning, we can leave the three benchmark datasets intact to enable a fair comparison with other papers. 
However, if there is a domain shift between the patches included in the fine-tuning and benchmark datasets, the resulting APR performance can be negatively affected.

\subsection{Models}

\noindent
In this section, we present the LLMs used for our experiments. 
We chose them based on their performance on the HumanEval dataset~\cite{chen2021:evaluating}, popularity and availability on Hugging Face, with the intention of improving over APR performance observed by Jiang et al.~\cite{jiang2023impact}. 
Moreover, we consider CodeGen and CodeT5 models to partially replicate the results in the related literature~\cite{jiang2023impact}.
DeepSeekCoder and CodeLlama-2 allow us to additionally explore the newer models and compare with the results of Li et al.~\cite{li2024:exploring}.
In total, we selected six different LLMs with different sizes:

\textbf{CodeGen}~\cite{nijkamp2023codegen} is a family of decoder models based on transformer architecture. Today there are 3 available versions: CodeGen-1, CodeGen-2, and CodeGen-2.5 from which CodeGen-1, or CodeGen, will be used. Models were pre-trained on the next token prediction task on three datasets: ThePile, BigQuery, and BigPython, resulting in CodeGen-NL, CodeGen-Multi, and CodeGen-Mono versions. In our experiment, we use CodeGen-Multi with three sizes: 350M, 2B to 6B. %

\textbf{CodeT5}~\cite{wang2021codet5} is a family of encoder-decoder models based on transformer architecture. Nowadays, there is a newer version is available, CodeT5+, but for our experiment, CodeT5 is used to replicate the model selection and results by Jiang et al.~\cite{jiang2023impact}. The models were pre-trained using the Masked Identifier Prediction task on the CodeSearchNet dataset, resulting in CodeT5-small (60M), CodeT5-base (220M) and CodeT5-large (770M) versions. %

\textbf{StarCoder}~\cite{li2023starcoder} is a family of decoder models based on transformer architecture. Models were pre-trained using the Fill-in-the-Middle (FIM) task on 80 programming languages to produce StarCoderBase and further fine-tuned on Python code to produce StarCoder. We are using StarCoderBase and its 1B and 3B versions. %

\textbf{DeepSeekCoder}~\cite{guo2024:deepseekcoder} is a family of transformer-based decoder models. Models were pre-trained using a Fill-in-the-Middle approach on the custom dataset consisting of 87\% code, 10\% code-related language and 3\% noncode-related Chinese language~\cite{guo2024:deepseekcoder} producing DeepSeekCoder-Base and further tuned through instruction-based data to produce DeepSeekCoder-Instruct. We use the 1.3B and 6.7B versions of DeepSeekCoder-Base. 
Note that at the time of writing, we used the latest DeepSeekCoder model available, i.e., v1.

\textbf{Bloom}~\cite{workshop2023bloom} is a family of decoder models based on transformer architecture. Models were pre-trained using next token prediction on the ROOTS dataset consisting of 46 natural and 13 programming languages. We use versions 560M, 1B7 (1B700M) and 7B1 (7B100M). %

\textbf{CodeLlama2}~\cite{roziere2024:codellama} is a family of decoder models based on transformer architecture built on the Llama2 family of models~\cite{touvron2023llama}. Models were pre-trained using Fill-in-the-Middle task on custom datasets starting from versions of Llama2 resulting in CodeLlama2 models, which were further tuned on Python to produce CodeLlama2-Python along with instruction-based dataset resulting in CodeLlama2-Instruct. We use the 7B version of CodeLlama2. %

\subsection{Evaluation}

\noindent
To measure the ability of LLMs to fix one of the benchmarking problems, we let them generate 10 patches for each problem, as done in related works~\cite{jiang2023impact,silva2024:repairllama,li2024comprehensive,yang2024multi,li2024:exploring}. 
Each patch is then evaluated with the unit tests for the corresponding problem. The results of the tests determine the success of the patch, more specifically they provide detailed information about the execution of the patch. The results of the patch execution can be either:
\begin{itemize}
    \item \textbf{Plausible (P):} Patch was compiled and the project successfully passed through all of its test(s).
    \item \textbf{Timeout (T):} Patch compilation and execution exceeded the time limit.
    \item \textbf{Uncompilable (U):} Patch was not compilable and the project, therefore, could not be further executed.
    \item \textbf{Wrong (W):} Patch was compiled successfully but the project did not pass its test(s).
    \item \textbf{Unknown (UNK):} Unique error due to the empty patch, due to the difference between the benchmarks and the Java version. This leads to the unavailability of execution for the generation of patches. 
\end{itemize}
For the proceeding experiments, we focus on plausible patches (i.e., we count if any of the 10 patches is plausible), as an indication of their correctness.

In addition to the performance on tests, we investigate the quality of generated patches based on their closeness to the ground truth, human patches. 
This is in particular helpful for patches that do not solve all the tests. %
To compute the similarity between two code snippets, we use CodeBLEU.
CodeBLEU~\cite{ren2020:codebleu} extends the BLEU metric~\cite{papineni-etal-2002-bleu} to be suitable for use in the quality assessment of code generation models. Similarly to BLEU, CodeBLEU is used for comparison between the generated code and the reference output. 
CodeBLEU takes into account not only the $n$-gram match as BLEU, but also considers the logic, syntax, and structure of the code. Therefore, CodeBLEU is often used in the evaluation of code generation models and is also used in many benchmarks for code, such as CodeXGLUE~\cite{lu2021codexglue}.

\subsection{Implementation Details}
\label{section:implementation}

\noindent
We use Hugging Face\footnote{~\url{https://huggingface.co/}} as the provider for the selected LLMs and implementations of the LoRA and IA3 PEFT techniques.\footnote{~\url{https://huggingface.co/docs/peft/index}}

To use datasets for inference and fine-tuning, the inputs need to be pre-processed for each model.
The corresponding pre-processing for each model is shown in Listing \ref{code:preproc_fine}, inspired by Jiang et al.~\cite{jiang2023impact} with the addition of FIM pre-processing.
In our experiments, we consider two fine-tuning scenarios: with buggy line, without buggy line.
The scenario ``with buggy line'' uses the prompts as show in Listing \ref{code:preproc_fine}, while the scenario ``without buggy line'' removes the lines ''// bug start'', ``// bug end'', and any line in between.
Bloom and CodeGEN use formatting similar to FIM, helping models to predict only replacement for the buggy line, instead of the code after the buggy line. CodeT5 models are fine-tuned in an encoder-decoder fashion, where the commented target is used only for the output. The remaining models are trained using the FIM approach~\cite{bavarian2022:efficient}.

\begin{listing}[b]
\centering
\begin{minipage}{.5\columnwidth}
\begin{minted}[fontsize=\scriptsize]{java}
// Bloom:
public static int 
       bitcount(int n) {
    int count = 0;
    while (n != 0) {
        // bug start:   
        n = (n ^ (n - 1));
        // bug end 
        count++;
    }
    return count;
}
// fix: 
n = (n & (n - 1));

\end{minted}
\end{minipage}\hfill
\begin{minipage}{.5\columnwidth}
\begin{minted}[fontsize=\scriptsize]{java}
// CodeGEN:
public static int 
       bitcount(int n) {
    int count = 0;
    while (n != 0) {
        // bug start: 
        n = (n ^ (n - 1));
        // bug end 
        count++;
    }
    return count;
}
// fix: 
n = (n & (n - 1));

\end{minted}
\end{minipage}

\begin{minipage}{.5\columnwidth}
\begin{minted}[fontsize=\scriptsize]{java}
// CodeLlama2:
public static int 
       bitcount(int n) {
    int count = 0;
    while (n != 0) {
        // bug start: 
        n = (n ^ (n - 1));
        // bug end 
        <FILL_ME>
        count++;
    }
    return count;
}
n = (n & (n - 1));

\end{minted}
\end{minipage}\hfill
\begin{minipage}{.5\columnwidth}
\begin{minted}[fontsize=\scriptsize]{java}
// CodeT5:
public static int 
       bitcount(int n) {
    int count = 0;
    while (n != 0) {
        // bug start: 
        n = (n ^ (n - 1));
        // bug end
        count++;
    }
    return count;
}
// n = (n & (n - 1));

\end{minted}
\end{minipage}

\begin{minipage}{.5\columnwidth}
\begin{minted}[fontsize=\scriptsize]{java}
// DeepSeekCoder:
<|fim_begin|>
public static int 
       bitcount(int n) {
    int count = 0;
    while (n != 0) {
        // bug start: 
        n = (n ^ (n - 1));
        // bug end 
        <|fim_hole|>
        count++;
    }
    return count;
}
<|fim_end|>
n = (n & (n - 1));
\end{minted}
\end{minipage}\hfill
\begin{minipage}{.5\columnwidth}
\begin{minted}[fontsize=\scriptsize]{java}
// StarCoder:
<fim_prefix>
public static int 
       bitcount(int n) {
    int count = 0;
    while (n != 0) {
        // bug start: 
        n = (n ^ (n - 1));
        // bug end
        <fim_suffix>
        count++;
    }
    return count;
}
<fim_middle>
n = (n & (n - 1));
\end{minted}
\end{minipage}
\caption{Illustration of input code pre-processing for fine-tuning of selected models with the last line being the target, correct fix for the buggy line.}
\label{code:preproc_fine}
\end{listing}

\section{Results and Discussion}

\noindent
In the first research question, we start by testing a set of models on three APR benchmarks without fine-tuning the models. The benchmarks consist of Java code and tests and vary in size and complexity of the bug. We proceed with model selection and fine-tuning on a dedicated APR dataset to study the performance of fine-tuned models on the same benchmarks (RQ2). To evaluate performance during fine-tuning, we use exact match and CodeBLEU. Finally, we used original models to compare the effect of full fine-tuning and parameter-efficient fine-tuning on APR results (RQ3). 
Given the closeness to the work by Li et al.~\cite{li2024:exploring}, we compare our results with theirs were possible.
These comparisons are done for the identical models and datasets, as well as 10 patches for evaluation.

\begin{table}
\centering
\caption{RQ1 (no fine-tuning): number of plausible patches for QuixBugs (QB), HumanEval-Java (HE), Defects4J (D4J) compared with Jiang et al.~\cite{jiang2023impact} in parentheses, with the best results for each model on each benchmark highlighted in bold.}
\label{tab:rq1}
\resizebox{\columnwidth}{!}{
\begin{tabular}{lcccaaa}
\toprule
\multirow{1}{*}{} &
  \multicolumn{3}{c}{Without buggy line} &
  \multicolumn{3}{c}{With buggy line} \\
\toprule
\rowcolor{white}  {Model} &  QB &  HE &  D4J &  QB &  HE &  D4J \\
\toprule
Bloom-560m &           1 &  \textbf{10} &  \textbf{8} &  \textbf{4} & 5 & 6 \\
Bloom-1b7  &           \textbf{1} &  \textbf{15} &  \textbf{10} & \textbf{1} & 12 & 8 \\
Bloom-7b1  &           \textbf{6} &  \textbf{24} &  10 & 4 & 22 & \textbf{14}\\
\midrule
CodeGen-350M &        \textbf{6}(7)  & \textbf{25}(30) & \textbf{12}(3) & 4  & 16 & 7 \\
CodeGen-2B   &       \textbf{13}(15) & \textbf{44}(49) & \textbf{20}(4) & 12 & 34 & \textbf{20} \\
CodeGen-6B   &       \textbf{16}(16) & \textbf{42}(46) & \textbf{18}(8) & \textbf{16} & \textbf{42} & \textbf{18} \\
\midrule
CodeLlama2-7b &      \textbf{33} & 95 & 83 & 28 & \textbf{96} & \textbf{93} \\
\midrule
CodeT5-small &      \textbf{0}(3) & \textbf{4}(3) & \textbf{10}(2) & \textbf{0} & 1 & 4 \\
CodeT5-base  &      \textbf{0}(0) & \textbf{4}(5) & \textbf{8}(4)  & \textbf{0} & 0 & 2 \\
CodeT5-large &      \textbf{2}(3) & 2(6) & \textbf{4}(1)  & 1 & \textbf{3} & \textbf{4} \\
\midrule
DeepSeekCoder-1.3b &   \textbf{33} & 94 &  72 & 31 & \textbf{97}  & \textbf{86}\\
DeepSeekCoder-6.7b  &  \textbf{33} & \textbf{107} & 89 & 30 & 105 & \textbf{103}\\
\midrule
StarCoder-1b &      \textbf{22} & 69 &  62 & 22 & \textbf{70} & \textbf{74} \\
StarCoder-3b &      \textbf{32} & \textbf{94} &  63 & 31 & 85 & \textbf{87} \\
StarCoder-7b &      \textbf{33} & \textbf{91} &  82 & 31 & \textbf{91} & \textbf{96}  \\
\bottomrule \\
\end{tabular}}
\end{table}

\subsection{RQ1: Performance without Fine-tuning}

\noindent
In the first research question, we investigate the performance of the 15 LLMs on the three APR benchmarking datasets (QuixBugs, HumanEval-Java, Defects4J) without any fine-tuning. 
In accordance with Jiang et al.~\cite{jiang2023impact}, we query the LLMs with and without highlighting the buggy lines.

Table~\ref{tab:rq1} summarizes the number of plausible patches achieved by each model for the three datasets.
We can observe that the number of model parameters effects model performance, where more parameters usually correspond to better performing models. 
One example for this is CodeGen, for which CodeGen-2B and CodeGen-6B always generate more plausible patches than CodeGen-350M. 
The same holds for StarCoder, where the smallest model, StarCoder-1b, generates the least plausible patches in all six cases. 
Once the models reach a size of a billion parameters, the differences are less pronounced, with 2B and 3B models able to tie or even generate more plausible patches than their 7B counterparts. 
For CodeT5 however, we observe the opposite. CodeT5-small has the best performance in 3 out of 6 cases. 

Moreover, the results compared to Jiang et al.~\cite{jiang2023impact} differ slightly for QuixBugs and HumanEval-Java and substantially on Defects4J benchmarks. 
There are multiple reasons: for instance, Jiang et al. did not specify the Java version used, furthermore, they filter out more programs based on their length, resulting in fewer fixes. 

Importantly, pre-training of models uses large datasets, usually mined from GitHub. This may cause data leakage, where the benchmarks may have been seen by models at some point during pre-training because they are available as public repositories on GitHub. This could explain the behavior of best-performing (newer) models. If we consider the total number of buggy programs with the number of fixed ones, without buggy line, by the best performing model DeepSeekCoder-6.7b we obtain 82.5\% of fixed codes for QuixBugs, with 65.6\% for HumanEval-Java and 40.6\% for Defects4J. %

Intuitively, by adding a buggy line, we could steer the model in the direction of a potential fix. Thus, the model does not have to come up with the solution fully, but can utilize the already existing, wrong solution.
However, as we can see in Table \ref{tab:rq1}, adding buggy lines did not improve performance in 29 out of 45 cases. %
Adding buggy lines for QuixBugs is rarely useful, it only increases the number of plausible patches for Bloom-560 from 1 to 4.
While adding a buggy line does not lead to performance improvements in the majority of cases, it does improve the best performing models (CodeLlama, DeepSeekCoder, StarCoder).
In particular, the performance on Defects4J is improved for all six models sizes, and in three out of six sizes for HumanEval. 
One reason for such an inconsistent effect of including buggy lines in the repair task is the pre-training phase of the models, which determines if and how they are able to make use of such additional information. 

Compared to the results of Li et al.~\cite{li2024:exploring} (Table~4, no fine-tuning), the CodeLlama2-7B model in our work solves more problems on all datasets. 
We appoint this difference to different prompting strategies and general stochasticity of LLM inference.
The trend of larger models performing better holds in the study of Li et al.~\cite{li2024:exploring} as well. 
Comparing DeepSeekCoder results without fine-tuning from Table~\ref{tab:rq1} is not possible because Li et al.~\cite{li2024:exploring} only reported results of the fine-tuned model with adapters.

\begin{framed}
    \noindent \textbf{To summarize the answer to RQ1:} Benchmarking performance varies between each model, where DeepSeekCoder-6.7b generates the largest number of plausible patches for QuixBugs (33 out of 40 problems), HumanEval-Java (107 out of 163 problems) and Defects4J (89 out of 219 problems). %
    Larger models perform better in 34 out of 48 cases, while the usefulness of adding buggy lines varies across models.
\end{framed}

\begin{table*}
\caption{RQ2 (full-model fine-tuning): Model performance on validation dataset (for epochs 0, 1, 2, and 3). Results on the \linebreak training dataset are displayed for epoch 3. The best results on the validation dataset are highlighted in bold. \linebreak The arrows after metrics indicate whether larger ($\uparrow$) or smaller ($\downarrow$) values are considered better.}
\label{tab:rq2_1}
\centering
\resizebox{.8\textwidth}{!}{
\setlength{\tabcolsep}{3pt}
\begin{tabular}{lrrrra@{\hspace{20pt}}rrrra@{\hspace{20pt}}rrrra}
\toprule
& \multicolumn{5}{c}{CodeBLEU ($\uparrow$)} & \multicolumn{5}{c}{Loss ($\downarrow$)} & \multicolumn{5}{c}{Exact Match ($\uparrow$)} \\  \midrule
\rowcolor{white}
 & \multicolumn{4}{c}{Validation Epoch} &  & \multicolumn{4}{c}{Validation Epoch} &  & \multicolumn{4}{c}{Validation Epoch} &  \\ 
\rowcolor{white} & 0 & 1 & 2 & 3 & Train & 0 & 1 & 2 & 3 & Train & 0 & 1 & 2 & 3 & Train \\ \midrule
Bloom-560m & 0.45 & 0.49 & 0.54 & 0.57 & 0.73 & 1.21 & 0.88 & 0.74 & 0.76 & 0.25 & 0.00 & 0.06 & 0.13 & 0.19 & 0.53 \\
Bloom-1b7 & 0.46 & 0.53 & 0.58 & 0.60 & 0.84 & 1.18 & 0.68 & 0.59 & 0.66 & 0.14 & 0.00 & 0.08 & 0.16 & 0.21 & 0.69 \\ \midrule
CodeGen-350M & 0.41 & 0.57 & 0.61 & 0.62 & 0.86 & 1.06 & 0.49 & 0.46 & 0.55 & 0.13 & 0.00 & 0.11 & 0.18 & 0.22 & 0.70 \\
CodeGen-2B & 0.38 & 0.53 & 0.59 & 0.61 & 0.88 & 0.90 & 0.58 & 0.52 & 0.64 & 0.13 & 0.00 & 0.10 & 0.19 & 0.23 & 0.79 \\ \midrule
CodeT5-small & 0.05 & 0.38 & 0.39 & 0.38 & 0.39 & 4.37 & 0.69 & 0.66 & 0.65 & 0.59 & 0.00 & 0.03 & 0.05 & 0.05 & 0.06 \\
CodeT5-base & 0.07 & 0.36 & 0.38 & 0.39 & 0.45 & 4.78 & 0.57 & 0.53 & 0.52 & 0.33 & 0.00 & 0.08 & 0.12 & 0.14 & 0.25 \\
CodeT5-large & 0.01 & 0.36 & 0.42 & 0.43 & 0.52 & 4.38 & 0.56 & 0.50 & 0.50 & 0.23 & 0.00 & 0.12 & 0.19 & 0.23 & 0.41 \\ \midrule
DeepSeekCoder-1.3b & 0.45 & 0.60 & 0.64 & \textbf{0.65} & 0.89 & 0.75 & \textbf{0.44} & \textbf{0.44} & 0.56 & 0.10 & 0.00 & 0.15 & 0.23 & \textbf{0.26} & 0.82 \\ \midrule
StarCoder-1b & 0.44 & 0.58 & 0.61 & 0.63 & 0.87 & 1.05 & 0.57 & 0.53 & 0.59 & 0.11 & 0.00 & 0.12 & 0.20 & 0.24 & 0.72 \\
StarCoder-3b & 0.46 & 0.52 & 0.60 & 0.62 & 0.86 & 0.89 & 0.64 & 0.58 & 0.66 & 0.13 & 0.00 & 0.11 & .019 & 0.24 & 0.75 \\ \bottomrule \\
\end{tabular}
}
\end{table*}

\subsection{RQ2: Performance with Fine-tuning}

\noindent
RQ1 illustrated the performance of LLMs without fine-tuning, and we observed that some models, such as CodeT5, have rarely produced plausible patches. 
In this RQ, we investigate whether the performance of the pre-trained LLMs can be improved by fine-tuning them for the APR task.
Therefore, we fully fine-tune selected models on the CLM dataset (Section~\ref{section:datasets}). 
Given the inconsistent results from RQ1, we decided to focus the following investigation on scenarios without including buggy lines for fine-tuning. 
In the work by Jiang et al.~\cite{jiang2023impact}, models were fine-tuned for only one epoch. The reason being computation time and the non-significant decrease in loss after the first epoch. 
Here, we fine-tune models for 3 epochs to see if there is additional improvement of the models after three epochs, and more importantly provide numerical evidence. 

The dataset was pre-processed to match the specific input format for each of the selected models (Section~\ref{section:implementation}).
Furthermore, the CLM dataset is split into training and validation datasets. We log the fine-tuning process of models after every epoch to see their performance change on the CLM validation dataset with loss, CodeBLEU and exact match. %

We selected models for fine-tuning based on their resource usage.
The deciding factor is the computational resource, which we chose to be 1 node with A100, and V100 GPUs, however, note that larger models can be trained on several GPUs. This led to fine-tuning of models with less than 3b parameters, resulting in the selection of Bloom-560m/1b7, CodeGen-350M/2B, CodeT5-small/base/large, DeepSeekCoder-1.3b, and StarCoder-1b/3b. 

\begin{table}[b]
\caption{RQ2 (full-model fine-tuning): number of plausible patches for QuixBugs (QB), HumanEval-Java (HE), and Defects4J (D4J). The best results for each model on each benchmark are highlighted in bold.}
\label{tab:rq2_2}
\centering
\resizebox{\columnwidth}{!}{
\setlength{\tabcolsep}{3pt}
\begin{tabular}{lrrcaaarrr}
\toprule
\multirow{1}{*}{} &
  \multicolumn{3}{c}{Base} &
  \multicolumn{3}{c}{Epoch 1} &
  \multicolumn{3}{c}{Epoch 3} \\
\toprule
\rowcolor{white} {Model} &  QB &  HE &  D4J &  QB &  HE &  D4J &  QB &  HE &  D4J \\
\toprule
Bloom-560m &   1 & 10 & 8 &        5 &  17 &  \textbf{41} & \textbf{6} & \textbf{19} & \textbf{41} \\
Bloom-1b7  &   1 & 15 & 10 &        8 &  10 &  43 & \textbf{11} & \textbf{23} & \textbf{53} \\
\midrule
CodeGen-350M & 6 & 25 & 12 &       13  & \textbf{43} & \textbf{67} & \textbf{16}  & 42 & 61 \\
CodeGen-2B   & \textbf{13} & \textbf{44} & 20 &      8 & 25 & \textbf{66} & 11 & 36 & 64 \\
\midrule
CodeT5-small & 0 & 4 & 10 &     \textbf{14} & \textbf{44} & 53 & 13 & 39 & \textbf{60} \\
CodeT5-base  & 0 & 4 & 8 &   14 & \textbf{45} & 68  & \textbf{17} & 39 & \textbf{75} \\
CodeT5-large & 2 & 2 & 4 &    9 & 36 & 62  & \textbf{16} & \textbf{47} & \textbf{72} \\
\midrule
DeepSeekCoder-1.3b & \textbf{33} & \textbf{94} & 72 & 19 & 54 & \textbf{84} & 15 & 64  & 80\\
\midrule
StarCoder-1b &  \textbf{22} & \textbf{69} & 62 &   12 & 41 &  \textbf{72} & 13 & 49 & 71 \\
StarCoder-3b &  \textbf{32} & \textbf{94} & \textbf{63} &   6 & 26 &  51 & 11 & 37 & \textbf{63} \\
\bottomrule \\
\end{tabular}}
\end{table}

Table~\ref{tab:rq2_1} lists the model behavior during the three training epochs.
As can be seen in Table, the models improve all three metrics (CodeBLEU, loss and exact match) on the validation dataset.
CodeT5 models improved the most, since they had the worst loss and CodeBLEU values at the start. The best performing model in Table \ref{tab:rq2_1} is DeepSeekCoder-1.3b, with a CodeBLEU of 0.65 and an exact match of 0.26 on the validation dataset after three epochs of fine-tuning. 
Notice the difference between CodeBLEU and exact match, where the model understands the code structure but makes small mistakes, leading to loss of exact match. 
Smaller CodeT5 models have a similar loss on training and validation datasets, from which we hypothesize that they do not overfit, while the remaining models overfit to a higher extent. %

All models seem unable to capture the complexity of program repair tasks as they are trained further. It is visible by considering the progress of losses as we train models further, where in some cases validation loss stagnates at high values or even increases at epoch 3. There may be multiple reasons, one of them being the high variability of the CLM dataset. Another potential reason is the use of relatively small models for complex tasks. However, compared to Jiang et al.~\cite{jiang2023impact}, we observe improvements in exact match and CodeBLEU even though the loss stagnates. Therefore, there is potential for training models beyond one epoch.

\begin{table*}
\centering
\caption{RQ3 (PEFT): Model performance on validation dataset (for epochs 0, 1, 2, and 3). Results on the training 
\linebreak dataset are displayed for epoch 3. The best results on the validation dataset are highlighted in bold.
\linebreak The arrows after metrics indicate whether larger ($\uparrow$) or smaller ($\downarrow$) values are considered better.}
\label{tab:rq3}
\resizebox{.8\textwidth}{!}{
\setlength{\tabcolsep}{3pt}
\begin{tabular}{lrrrra@{\hspace{20pt}}rrrra@{\hspace{20pt}}rrrra}
\toprule
& \multicolumn{5}{c}{CodeBLEU ($\uparrow$)} & \multicolumn{5}{c}{Loss ($\downarrow$)} & \multicolumn{5}{c}{Exact Match ($\uparrow$)} \\  \midrule
\rowcolor{white}
 & \multicolumn{4}{c}{Validation Epoch} &  & \multicolumn{4}{c}{Validation Epoch} &  & \multicolumn{4}{c}{Validation Epoch} &  \\ 
 \rowcolor{white} & 0 & 1 & 2 & 3 & Train & 0 & 1 & 2 & 3 & Train & 0 & 1 & 2 & 3 & Train \\ \midrule
\rowcolor{lightgray}  \multicolumn{16}{c}{LoRA} \\ \midrule
CodeGen-350M & 0.39 & 0.58 & 0.58 & 0.58 & 0.60 & 1.12 & 0.47 & 0.47 & 0.46 & 0.44 & 0.00 & 0.06 & 0.07 & 0.08 & 0.09 \\
CodeGen-2B & 0.35 & 0.61 & 0.62 & 0.62 & 0.64 & 0.94 & 0.40 & 0.39 & 0.39 & 0.34 & 0.00 & 0.10 & 0.13 & 0.13 & 0.17 \\
CodeGen-6B & 0.42 & 0.61 & 0.62 & 0.62 & 0.65 & 0.93 & 0.38 & 0.38 & \textbf{0.37} & 0.33 & 0.00 & 0.11 & \textbf{0.14} & \textbf{0.14} & 0.20 \\ \midrule
CodeT5-small & 0.04 & 0.33 & 0.35 & 0.34 & 0.34 & 4.39 & 0.84 & 0.82 & 0.81 & 0.79 & 0.00 & 0.02 & 0.03 & 0.03 & 0.03 \\
CodeT5-base & 0.08 & 0.34 & 0.35 & 0.34 & 0.35 & 4.80 & 0.69 & 0.67 & 0.67 & 0.65 & 0.00 & 0.04 & 0.05 & 0.05 & 0.05 \\
CodeT5-large & 0.01 & 0.33 & 0.35 & 0.35 & 0.36 & 4.37 & 0.57 & 0.56 & 0.55 & 0.53 & 0.00 & 0.06 & 0.07 & 0.07 & 0.08 \\ \midrule
DeepSeekCoder-1.3b & 0.44 & 0.61 & 0.61 & 0.61 & 0.63 & 0.76 & 0.43 & 0.42 & 0.42 & 0.39 & 0.00 & 0.09 & 0.11 & 0.12 & 0.14 \\
DeepSeekC.-6.7b & 0.48 & \textbf{0.63} & \textbf{0.63} & \textbf{0.63} & 0.65 & 0.69 & 0.39 & 0.38 & 0.38 & 0.35 & 0.00 & 0.13 & \textbf{0.14} & \textbf{0.14} & 0.16 \\ \midrule
\rowcolor{lightgray} \multicolumn{16}{c}{IA3} \\ \midrule
CodeGen-350M & 0.41 & 0.56 & 0.56 & 0.56 & 0.57 & 1.08 & 0.53 & 0.52 & 0.52 & 0.51 & 0.00 & 0.02 & 0.02 & 0.02 & 0.03 \\
CodeGen-2B & 0.38 & 0.59 & 0.59 & 0.59 & 0.60 & 0.93 & 0.46 & 0.45 & 0.45 & 0.44 & 0.00 & 0.04 & 0.04 & 0.04 & 0.05 \\
CodeGen-6B & 0.43 & 0.59 & 0.59 & 0.59 & 0.60 & 1.08 & 0.48 & 0.47 & 0.46 & 0.44 & 0.00 & 0.04 & 0.05 & 0.05 & 0.05 \\ \midrule
CodeT5-small & 0.04 & 0.22 & 0.24 & 0.25 & 0.25 & 4.39 & 1.20 & 1.11 & 1.09 & 1.08 & 0.00 & 0.01 & 0.01 & 0.01 & 0.01 \\
CodeT5-base & 0.08 & 0.26 & 0.27 & 0.28 & 0.28 & 4.76 & 0.94 & 0.88 & 0.87 & 0.86 & 0.00 & 0.01 & 0.01 & 0.02 & 0.02 \\
CodeT5-large & 0.01 & 0.29 & 0.30 & 0.30 & 0.31 & 4.39 & 0.79 & 0.67 & 0.66 & 0.65 & 0.00 & 0.03 & 0.04 & 0.04 & 0.04 \\ \midrule
DeepSeekCoder-1.3b & 0.46 & 0.60 & 0.60 & 0.60 & 0.61 & 0.83 & 0.49 & 0.48 & 0.47 & 0.47 & 0.00 & 0.06 & 0.06 & 0.06 & 0.06 \\
DeepSeekC.-6.7b & 0.50 & \textbf{0.62} & \textbf{0.62} & \textbf{0.62} & 0.62 & 0.75 & 0.43 & \textbf{0.42} & \textbf{0.42} & 0.42 & 0.00 & 0.08 & 0.08 & \textbf{0.09 }& 0.09 \\ 
\bottomrule
\end{tabular}
}
\end{table*}

Table \ref{tab:rq2_2} shows the results of various models and their performance in terms of plausible repairs of programs without a buggy line, compared to the results from Table \ref{tab:rq1}.
From this table, we observe that the models, compared to the base (non fine-tuned), perform differently for each model. CodeT5, Bloom, and CodeGen show improvements, whereas DeepSeekCoder and StarCoder show deterioration after fine-tuning compared to using the base models. 

We argue that the reason for this deterioration is due to different data distributions. We used the CLM dataset, which may not be representative of various bugs encountered in the benchmarking datasets, including Defects4J, HumanEval-Java, and Quixbugs. One solution to improve the quality of models is to use larger fine-tuning datasets. 

\begin{framed}
    \noindent \textbf{To summarize the answer to RQ2:} Full fine-tuning leads to improvement of models that performed poorly without fine-tuning, such as CodeT5 and Bloom, but leads to worsening of the best performing models, including DeepSeekCoder. %
\end{framed}

\subsection{RQ3: Performance with Parameter-efficient Fine-tuning}

\noindent
In addition to standard full-model fine-tuning performed in RQ2, RQ3 investigates PEFT methods. 
In particular, we use LoRA and IA3. 
Furthermore, we investigate the impact of hyperparameters for fine-tuning with LoRA (RQ4).
Due to computational resources and time constraints, CodeGen, CodeT5, and DeepSeekCoder models were selected based on their performance. %

Similar to RQ2, we fine-tune the models for 3 epochs. 
We follow the default parameters recommended by Hugging Face PEFT,\footnote{~\url{https://huggingface.co/docs/peft/index}} where the default recommended rank and scaling factor of LoRA were $r=8, \alpha=16$ at the time of writing.
We follow Hugging Face default settings for IA3, too.%

Table~\ref{tab:rq3} shows the validation and training performance of the fine-tuned models using LoRA and IA3. 
Both methods improve the losses mainly up to the first epoch, where CodeBLEU and validation set loss stagnate for almost all the models. %
A possible reason for this is that adapters significantly reduce the number of trainable parameters and the PEFT-trained models reach their plateau performance at this point. %
Interestingly, Li et al.~\cite{li2024:exploring} noticed a drop in metrics with the increasing number of epochs (Table~6~\cite{li2024:exploring}), which goes in line with the finding that PEFT training does not necessarily benefit from more epochs.

We notice a large drop in the exact match compared to the full-model fine-tuning (Table~\ref{tab:rq2_1}). 
However, the drop in CodeBLEU is not as large: we observe rather high values with DeepSeekCoder-6.7b having a CodeBLEU of 0.63 (LoRA) and 0.62 (IA3). 
This shows that models can capture code structure similarly to full-finetuning, with the main difference being small mistakes like variable names, leading to low exact match. 
When using PEFT, models perform similarly for training and validation datasets, thus reducing overfitting. 
Finally, we point out the stagnation in metrics and loss in most models after the first epoch. For this reason, we chose the first epoch of each model fine-tuned using LoRA and IA3 and summarize their performance in Table~\ref{tab:rq3_bench}. 

In Table~\ref{tab:rq3_bench}, we can observe performance improvements with PEFT techniques over base models, in particular with LoRA on CodeGen and DeepSeekCoder. 
CodeGen with LoRA improves over full fine-tuning, especially on the Defects4J benchmark with 34 additional plausible patches on CodeGen-2b. 
In general, by using adapters we were able to achieve improvements on all benchmarks. 
Contrary to expectation and trends observed by Li et al.~\cite{li2024:exploring}, LoRA performs better than IA3 in 21 out of 24 cases. 
Li et al. reported that IA3 outperformed LoRA for DeepSeekCoder-6.7b, two CodeLlama model sizes but not Llama2-7b (see Table~5 in their study~\cite{li2024:exploring}). 
However, in the majority of our experiments, LoRA showed better performance than IA3, except for DeepSeekCoder-6.7b on D4J.
Furthermore, CodeT5 models improved even more compared to full-model fine-tuning on HumanEval-Java and Defects4J. 
We conclude that adapters are a viable approach to fine-tuning LLMs that offer, in most cases, additional improvements in terms of the number of plausible patches generated in addition to using less resources.

Lastly, we present the number of trainable parameters utilized during fine-tuning with LoRA and IA3 in Table~\ref{tab:parameters}. 
When LoRA and IA3 are used, models have $<1\%$ of the original number of trainable parameters. 
Importantly, we were able to fine-tune and fit larger original models (ca.~6B parameters) into GPU memory with the reduced number of parameters with respect to the PEFT method.

\begin{table}[t]
\centering
\caption{Number of plausible patches for QuixBugs (QB), HumanEval-Java (HE), and Defects4J (D4J) based on full-model fine-tuning (FMFT) for three epochs, PEFT using LoRA, IA3 trained for one epoch and compared with the base models without fine-tuning. The best results are highlighted in bold for each benchmark, model, and method. ``X'' indicates model configurations we were unable to train due to hardware limitations.}
\label{tab:rq3_bench}
\resizebox{\columnwidth}{!}{
\setlength{\tabcolsep}{2pt}
\begin{tabular}{lrrraaarrraaa}
\toprule
\multirow{1}{*}{} &
  \multicolumn{3}{c}{Base} &
  \multicolumn{3}{c}{FMFT} &
  \multicolumn{3}{c}{LoRA} &
  \multicolumn{3}{c}{IA3} \\
\toprule
{Model} &  QB &  HE &  D4J &  QB &  HE &  D4J &  QB &  HE &  D4J &  QB &  HE &  D4J \\
\toprule
CodeGen-350M &  6 & 25 & 12 &  \textbf{16}  & 42 & 61 & \textbf{16}  & \textbf{64} & \textbf{82} & 13  & 52 & 68 \\
CodeGen-2B   &  13 & 44 & 20 &  11 & 36 & 64 & \textbf{19} & \textbf{81} & \textbf{98} & 15 & 67 & 74 \\
CodeGen-6B   &  16 & 42 & 18 &  X &  X &  X & \textbf{24} & \textbf{81} & \textbf{104} & 14 & 70 & 83 \\
\midrule
CodeT5-small &  0 & 4 & 10 &  \textbf{13} & \textbf{39} & \textbf{60} & 10 & 36 & 50 & 5 & 18 & 42 \\
CodeT5-base  &  0 & 4 & 8 &  \textbf{17} & 39 & \textbf{75}  & 12 & \textbf{50} & 67  & 8 & 32 & 52 \\
CodeT5-large &  2 & 2 & 4 &  16 & 47 & \textbf{72} & \textbf{17} & \textbf{60} & \textbf{72}  & 10 & 47 & 61 \\
\midrule
DeepSeekCoder-1.3b & \textbf{33} & 94 & 72 &  15 & 64  & 80  & 27 & \textbf{106} & 100 & 28 & 94  & \textbf{101}\\
DeepSeekCoder-6.7b & \textbf{33} & 107 & 89 &  X &  X &  X  & 31 & \textbf{108} & 108 & 30 & 100 & \textbf{114}\\
\bottomrule \\
\end{tabular}}
\vspace*{-2ex}
\end{table}

\begin{framed}
\noindent \textbf{To summarize the answer to RQ3:} Fine-tuning in a parameter-efficient way with adapters leads to an improvement of several models compared to full-model fine-tuning, e.g., for CodeGen. For instance, by using LoRA for CodeGen-2B, we used only 0.09\% of trainable parameters of the full model, while achieving performance gains of 172\%, 225\%, 153\% on the QuixBugs, HumanEval-Java and Defects4J benchmarks. Furthermore, we observed that most of the time, LoRA achieves better results than IA3, specifically, in 21 out of 24 cases.
\end{framed}

\begin{table}[t]
\centering
\caption{Amount of trainable parameters for each of the used models with LoRA and IA3 adapter.}
\label{tab:parameters}
\resizebox{\columnwidth}{!}{
\begin{tabular}{lrrr}
\toprule
 & & 
\multicolumn{1}{c}{LoRA} & 
\multicolumn{1}{c}{IA3}\\
Model & Total & Trainable (\%) & Trainable (\%) \\
\toprule
CodeT5-small & 60,787,200  & 294,912 (0.49) & 43,008 (0.07)  \\
CodeT5-base & 223,766,784   & 884,736 (0.39) & 129,024 (0.06)  \\
CodeT5-large & 739,998,720  & 2,359,296 (0.32) & 344,064 (0.05)  \\
\midrule
CodeGen-350M & 357,367,808 & 655,360 (0.18)     & 61,440 (0.02)  \\
CodeGen-2B & 2,779,683,840 & 2,621,440 (0.09)    & 245,760 (9e-3) \\
CodeGen-6B & 7,068,538,880 & 4,325,376 (0.06)    & 405,504 (6e-3)  \\
\midrule
DeepSeekCoder-1.3b & 1,348,044,800 & 1,572,864 (0.12) & 230,400 (0.02) \\
DeepSeekCoder-6.7b & 6,744,707,072 & 4,194,304 (0.06)    & 614,400 (9e-3)\\
\bottomrule \\
\end{tabular}}
\end{table}

\subsection{RQ4: Effects of LoRA Parameters}

\noindent
LoRA keeps the parameters of pre-trained models frozen and adds a layer with a few trainable parameters. 
The main hyperparameters in LoRA that impact the resulting model size and also performance are rank and scaling factor (Section~\ref{section:lora}). 
Rank affects the amount of trainable parameters, whereas the scaling factor controls the magnitude of parameter updates. 
To test the effect of LoRA parameters on fine-tuning results, we performed experiments with several values of rank and scaling factor.
In particular, we investigate the following 8 values for both rank and scaling factor: 1, 2, 4, 8, 16, 32, 64. 
The default values for rank and scaling factor of LoRA are $r=8, \alpha=16$. 
We fix a default value for one and change the values of another hyperparameter in our experiments. 

\begin{figure}[tb]
    \centering
    \begin{minipage}{\columnwidth}
        \centering
        \includegraphics[width=\linewidth, trim=0 0 1.5cm 0, clip]{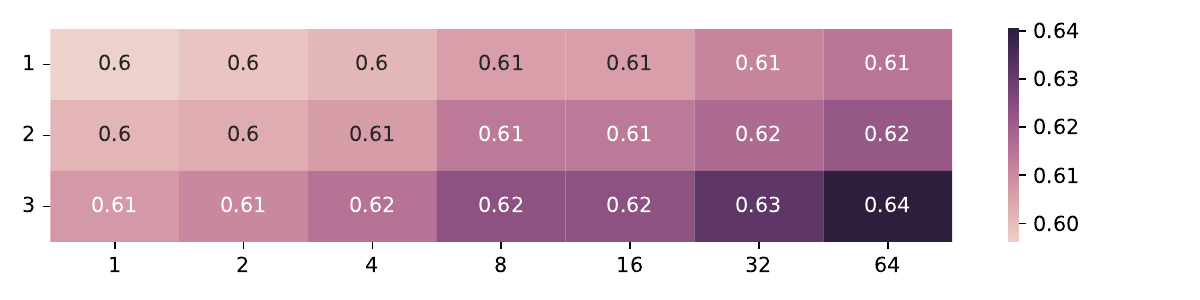}
        \vspace{-15pt}
        \caption{CodeBLEU change with scaling factor for three epochs.}
        \label{fig:scale-codebleu}
    \end{minipage}%

\vspace{15pt}

    \begin{minipage}{\columnwidth}
        \centering
        \includegraphics[width=\linewidth, trim=0 0 1.5cm 0, clip]{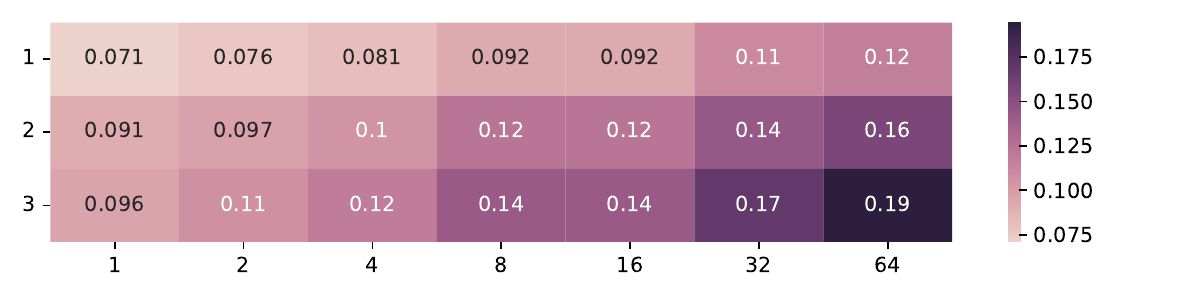}
        \vspace{-15pt}
        \caption{Exact match change with scaling factor for three epochs.}
        \label{fig:scale-exact}
    \end{minipage}%
    
\vspace{15pt}
    
    \begin{minipage}{\columnwidth}
        \centering
        \includegraphics[width=\linewidth, trim=0 0 1.5cm 0, clip]{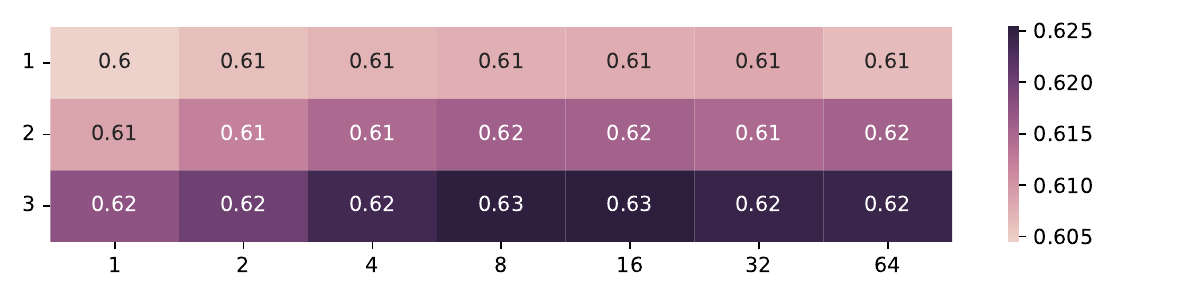}
        \vspace{-15pt}
        \caption{CodeBLEU change with rank for three epochs.}
        \label{fig:rank-codebleu}
    \end{minipage}
    
\vspace{15pt}
    
    \begin{minipage}{\columnwidth}
        \centering
        \includegraphics[width=\linewidth, trim=0 0 1.5cm 0, clip]{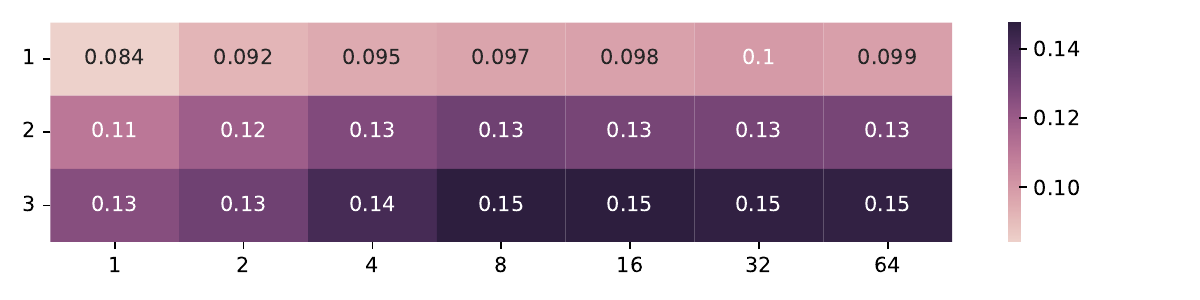}
        \vspace{-15pt}
        \caption{Exact match change with rank for three epochs.}
        \label{fig:rank-exact}
    \end{minipage}
\end{figure}

Figures \ref{fig:scale-codebleu} and \ref{fig:scale-exact} show the effect of the scaling factor while Figures \ref{fig:rank-codebleu} and \ref{fig:rank-exact} show the effect of the rank for LoRA when applied to CodeGen-2B. 
CodeBLEU is almost not affected by the change in rank or scaling factor and changes in the range between 0.6 and 0.64. 
However, the exact match shows larger deviations, interestingly more with the scaling factor than with rank. 
The conclusion is, however, the same: either rank or scaling factor does not affect the performance of models significantly. 

We observe that the exact match metric is affected to a slightly larger degree by the scaling factor than by rank size and stays rather low in all cases. 
However, because exact match evaluates the output against only one version of bug fix, this metric is not the best to optimize LLMs for. 
These results obtained agree with~\cite{lora}, where the change in LoRA hyperparameters does not play a significant role.

Li et al.~\cite{li2024:exploring} varied the rank hyperparameter for one model, CodeLlama-7b, and did not change the scaling factor. 
The study shows that the best rank in terms of fixing bugs on the APR datasets is 16 while the scaling factor was fixed to 16.
However, the study changes the rank while performing PEFT on a bug-fixing APR instructions dataset and testing the tuned models on the APR dataset. 
In our study, we notice the increase of the loss metrics with the increase of rank because we test LoRA on the validation part of the dataset used for PEFT, and observe a predictable increase in metrics values with the increase of the rank (and increase in trainable parameters). 
Intuitively, the more parameters are updated the better the metrics are on the data with the same distribution.

\begin{framed}
    \noindent \textbf{To summarize the answer to RQ4:} Both LoRA hyperparameters, scaling factor and rank, lead to slight, negligible differences in the resulting metrics during training. Thus, following the recommended hyperparameter values as compared to their search does not drastically affect the performance of models.
\end{framed}

\subsection{Threats to Validity}

\noindent
There are several factors that can influence the results of our experiments and their validity. 
First, benchmarks like HumanEval-Java and QuixBugs were created from simple projects and consist of bugs that are not representative of complex real-world bugs. Another issue arises in the validation of the Defects4J benchmark, due to the specific packages and Java versions used (i.e., for running the running and evaluating the patches), causing differences in the number of problems and results compared to those of Jiang et al.~\cite{jiang2023impact}. %

Additionally, versioning of various libraries caused by internal dependencies leads to internal problems of benchmarks, where some tests are not possible to execute.  Furthermore, the training APR dataset used that contains bug-fix pairs is not itself representative of bugs found in the benchmarks and has inherently different data distributions than the benchmarks themselves. To represent real-world complex bugs through benchmarks and training datasets, we would have to use much larger and more complex datasets.

The biggest reason for concern goes to data leakage, which can occur when models are pre-trained on a dataset on which we then test, skewing model performance~\cite{xu2024benchmark}. 
LLMs nowadays use very large datasets that sometimes are not publicly available or too large to even check whether they include tested benchmarks. 
None of the models achieved close to the 100\% benchmark performance, making it clear that the effect of data leakage is not the only factor that influences the performance of the model.

Lastly, we used plausibility to evaluate patches. 
Plausibility shows whether a patch passes all available tests but is not a guarantee of its correctness. 
One way to assess the correctness of patches is by manually checking.
However, manual correctness verification is prone to subjectivity~\cite{wang2020:automated}, 
which is why we remained with plausibility for patch evaluation.

\section{Conclusions and Future Work}

\noindent
In this work, we conduct an extensive empirical study on the impact of fine-tuning LLMs on their APR performance. 
We provide a detailed comparison of different fine-tuning strategies for APR and offer practical insights into leveraging LLMs for automated software maintenance and evolution.

We studied a total of six state-of-the-art LLMs that were pre-trained on code, 
and evaluated their bug fixing performance in three settings: 
without fine-tuning, with full-model fine-tuning, with parameter-efficient fine-tuning.
Our study has been inspired by the work of Jiang et al.~\cite{jiang2023impact} and draws parallels to the work of Li et al.~\cite{li2024:exploring}. 
We compared the models based on three popular APR datasets (QuixBugs, HumanEval-Java, and Defects4J).
Furthermore, buggy lines were included or omitted from the input to analyze whether additional information 
improves the LLM's performance before fine-tuning. 

Our findings show that some of the models decrease performance after 
full-model fine-tuning
in comparison to their non-tuned counterparts. 
Reasons for this can be overfitting, dataset variability, and size. 
Full fine-tuning led to the improvement of smaller models 
and 
the deterioration of APR performance on selected benchmarks for larger models compared to previously made inferences. To alleviate issues of full fine-tuning, such as overfitting and computational efficiency, we investigate two parameter-efficient fine-tuning techniques (LoRA and IA3). 
Among these two, we observed that LoRA achieved better results.
By using adapters as additional layers on top of frozen pre-trained models, we were able to solve issues of full-model fine-tuning and improve the performance of models compared to zero-shot inference for larger models. 

\head{Future work:}
Further research directions include the exploration of different adapters along with their interpretation. In addition, instruction-based models could provide additional help in steering the models toward a correct solution via adapting prompts and seem to be an interesting direction. 

The exploration of adapters for larger models would provide additional performance and computational insights. One of the questions worth answering is whether we can automate the selection of adapters, based on dataset and model, to achieve the best performance. Finally, the use of different optimizers would allow us to decrease the memory usage of models and train larger models, providing additional direction for the optimization and training of large models. 

\section{Data Availability}

\noindent
To support open science and allow for replication and verification of our work, a replication package with code and results is made available via Zenodo~\cite{machacek2025:replication}.

\section*{Acknowledgments}

This work is supported by the Research Council of Norway through the secureIT project (IKTPLUSS \#288787), and by the European Union through the Horizon Europe Marie Sk\l{}odowska-Curie Actions (\#101151798).
The empirical evaluation made use of the Experimental Infrastructure for Exploration of Exascale Computing (eX3), 
financially supported by the Research Council of Norway under contract \#270053. 

\balance
\printbibliography

\end{document}